\documentstyle[twoside,psfig]{article}

\newcommand{\bee}{\begin{equation}}
\newcommand{\eeq}{\end{equation}}

\newcommand{\bea}{\begin{eqnarray}}
\newcommand{\eea}{\end{eqnarray}}

\catcode`\@=11
\long\def\@makefntext#1{
\protect\noindent \hbox to 3.2pt {\hskip-.9pt  
$^{{\eightrm\@thefnmark}}$\hfil}#1\hfill}		%CAN BE USED 

\def\thefootnote{\fnsymbol{footnote}}
\def\@makefnmark{\hbox to 0pt{$^{\@thefnmark}$\hss}}	%ORIGINAL 
	
\def\ps@myheadings{\let\@mkboth\@gobbletwo
\def\@oddhead{\hbox{}
\rightmark\hfil\eightrm\thepage}   
\def\@oddfoot{}\def\@evenhead{\eightrm\thepage\hfil
\leftmark\hbox{}}\def\@evenfoot{}
\def\sectionmark##1{}\def\subsectionmark##1{}}

\oddsidemargin=\evensidemargin
\renewcommand{\thefootnote}{\fnsymbol{footnote}}

\newcounter{sectionc}\newcounter{subsectionc}\newcounter{subsubsectionc}
\renewcommand{\section}[1] {\vspace{12pt}\addtocounter{sectionc}{1} 
\setcounter{subsectionc}{0}\setcounter{subsubsectionc}{0}\noindent 
	{\tenbf\thesectionc. #1}\par\vspace{5pt}}
\renewcommand{\subsection}[1] {\vspace{12pt}\addtocounter{subsectionc}{1} 
	\setcounter{subsubsectionc}{0}\noindent 
	{\bf\thesectionc.\thesubsectionc. {\kern1pt \bfit #1}}\par\vspace{5pt}}
\renewcommand{\subsubsection}[1] {\vspace{12pt}\addtocounter{subsubsectionc}{1}
	\noindent{\tenrm\thesectionc.\thesubsectionc.\thesubsubsectionc.
	{\kern1pt \tenit #1}}\par\vspace{5pt}}
\newcommand{\nonumsection}[1] {\vspace{12pt}\noindent{\tenbf #1}
	\par\vspace{5pt}}

\newcounter{appendixc}
\newcounter{subappendixc}[appendixc]
\newcounter{subsubappendixc}[subappendixc]
\renewcommand{\thesubappendixc}{\Alph{appendixc}.\arabic{subappendixc}}
\renewcommand{\thesubsubappendixc}
	{\Alph{appendixc}.\arabic{subappendixc}.\arabic{subsubappendixc}}

\renewcommand{\appendix}[1] {\vspace{12pt}
        \refstepcounter{appendixc}
        \setcounter{figure}{0}
        \setcounter{table}{0}
        \setcounter{lemma}{0}
        \setcounter{theorem}{0}
        \setcounter{corollary}{0}
        \setcounter{definition}{0}
        \setcounter{equation}{0}
        \renewcommand{\thefigure}{\Alph{appendixc}.\arabic{figure}}
        \renewcommand{\thetable}{\Alph{appendixc}.\arabic{table}}
        \renewcommand{\theappendixc}{\Alph{appendixc}}
        \renewcommand{\thelemma}{\Alph{appendixc}.\arabic{lemma}}
        \renewcommand{\thetheorem}{\Alph{appendixc}.\arabic{theorem}}
        \renewcommand{\thedefinition}{\Alph{appendixc}.\arabic{definition}}
        \renewcommand{\thecorollary}{\Alph{appendixc}.\arabic{corollary}}
        \renewcommand{\theequation}{\Alph{appendixc}.\arabic{equation}}
        \noindent{\tenbf Appendix \theappendixc #1}\par\vspace{5pt}}
\newcommand{\subappendix}[1] {\vspace{12pt}
        \refstepcounter{subappendixc}
        \noindent{\bf Appendix \thesubappendixc. {\kern1pt \bfit #1}}
	\par\vspace{5pt}}
\newcommand{\subsubappendix}[1] {\vspace{12pt}
        \refstepcounter{subsubappendixc}
        \noindent{\rm Appendix \thesubsubappendixc. {\kern1pt \tenit #1}}
	\par\vspace{5pt}}

\newcommand{\textlineskip}{\baselineskip=13pt}
\newcommand{\smalllineskip}{\baselineskip=10pt}

\def\eightcirc{
\begin{picture}(0,0)
\put(4.4,1.8){\circle{6.5}}
\end{picture}}
\def\eightcopyright{\eightcirc\kern2.7pt\hbox{\eightrm c}} 

\newcommand{\copyrightheading}[1]
	{\vspace*{-2.5cm}\smalllineskip{\flushleft
	{\footnotesize Modern Physics Letters A, #1}\\
	{\footnotesize $\eightcopyright$\, World Scientific Publishing
	 Company}\\
	 }}

\newcommand{\publisher}[2]{{\begin{center}\footnotesize\smalllineskip 
	Received #1\\
	Revised #2
	\end{center}
	}}

\def\abstracts#1#2#3{{
	\centering{\begin{minipage}{4.5in}\footnotesize\baselineskip=10pt
	\parindent=0pt #1\par 
	\parindent=15pt #2\par
	\parindent=15pt #3
	\end{minipage}}\par}} 

\newcommand{\bibit}{\nineit}
\newcommand{\bibbf}{\ninebf}
\renewenvironment{thebibliography}[1]
	{\frenchspacing
	 \ninerm\baselineskip=11pt
	 \begin{list}{\arabic{enumi}.}
        {\usecounter{enumi}\setlength{\parsep}{0pt}     
	 \setlength{\leftmargin 12.7pt}{\rightmargin 0pt} %FOR 1--9 ITEMS
         \setlength{\itemsep}{0pt} \settowidth
	{\labelwidth}{#1.}\sloppy}}{\end{list}}

\newcounter{itemlistc}
\newcounter{romanlistc}
\newcounter{alphlistc}
\newcounter{arabiclistc}

\newcommand{\fcaption}[1]{
        \refstepcounter{figure}
        \setbox\@tempboxa = \hbox{\footnotesize Fig.~\thefigure. #1}
        \ifdim \wd\@tempboxa > 5in
           {\begin{center}
        \parbox{5in}{\footnotesize\smalllineskip Fig.~\thefigure. #1}
            \end{center}}
        \else
             {\begin{center}
             {\footnotesize Fig.~\thefigure. #1}
              \end{center}}
        \fi}

\newcommand{\tcaption}[1]{
        \refstepcounter{table}
        \setbox\@tempboxa = \hbox{\footnotesize Table~\thetable. #1}
        \ifdim \wd\@tempboxa > 5in
           {\begin{center}
        \parbox{5in}{\footnotesize\smalllineskip Table~\thetable. #1}
            \end{center}}
        \else
             {\begin{center}
             {\footnotesize Table~\thetable. #1}
              \end{center}}
        \fi}

\def\@citex[#1]#2{\if@filesw\immediate\write\@auxout
	{\string\citation{#2}}\fi
\def\@citea{}\@cite{\@for\@citeb:=#2\do
	{\@citea\def\@citea{,}\@ifundefined
	{b@\@citeb}{{\bf ?}\@warning
	{Citation `\@citeb' on page \thepage \space undefined}}
	{\csname b@\@citeb\endcsname}}}{#1}}

\newif\if@cghi
\def\cite{\@cghitrue\@ifnextchar [{\@tempswatrue
	\@citex}{\@tempswafalse\@citex[]}}
\def\citelow{\@cghifalse\@ifnextchar [{\@tempswatrue
	\@citex}{\@tempswafalse\@citex[]}}
\def\@cite#1#2{{$\null^{#1}$\if@tempswa\typeout
	{IJCGA warning: optional citation argument 
	ignored: `#2'} \fi}}

\def\pmb#1{\setbox0=\hbox{#1}
	\kern-.025em\copy0\kern-\wd0
	\kern.05em\copy0\kern-\wd0
	\kern-.025em\raise.0433em\box0}

\def\fnt#1#2{\footnotetext{\kern-.3em
	{$^{\mbox{\scriptsize #1}}$}{#2}}}

\def\fpage#1{\begingroup
\voffset=.3in
\thispagestyle{empty}\begin{table}[b]\centerline{\footnotesize #1}
	\end{table}\endgroup}

\def\runninghead#1#2{\pagestyle{myheadings}
\markboth{{\protect\footnotesize\it{\quad #1}}\hfill}
{\hfill{\protect\footnotesize\it{#2\quad}}}}
\font\tenrm=cmr10
\font\tenit=cmti10 
\font\tenbf=cmbx10
\font\bfit=cmbxti10 at 10pt
\font\ninerm=cmr9
\font\nineit=cmti9
\font\ninebf=cmbx9
\font\eightrm=cmr8

\def\qed{\hbox{${\vcenter{\vbox{			%HOLLOW SQUARE
   \hrule height 0.4pt\hbox{\vrule width 0.4pt height 6pt
   \kern5pt\vrule width 0.4pt}\hrule height 0.4pt}}}$}}

\renewcommand{\thefootnote}{\fnsymbol{footnote}}	%USE SYMBOLIC FOOTNOTE

\begin{document}
\setlength{\textheight}{7.7truein}  %for 2nd page onwards

\runninghead{Thermodynamics and/or of horizon: A unified approach $\ldots$}{Thermodynamics and/or of horizon: A unified approach  $\ldots$}

\normalsize\textlineskip
\thispagestyle{empty}
\setcounter{page}{1}

\copyrightheading{}			%{Vol. 0, No.0 (1992) 000--000}

\vspace*{0.88truein}

\fpage{1}
\centerline{\bf THERMODYNAMICS AND/OF HORIZONS: }
\centerline{\bf A COMPARISON OF SCHWARZSCHILD,}
\centerline{\bf  RINDLER AND deSITTER SPACETIMES}
\baselineskip=13pt
%\centerline{\bf MANUSCRIPTS USING COMPUTER SOFTWARE\footnote{For
 %the title, try not to use more than 3 lines. Typeset the title
%in 10 pt Times Roman, uppercase and boldface.}}
\vspace*{0.37truein}
\centerline{\footnotesize T. PADMANABHAN}
\baselineskip=12pt
\centerline{\footnotesize\it IUCAA, P.O.Box 4, Ganeshkhind,}
\baselineskip=10pt
\centerline{\footnotesize\it Pune 411 007, Maharashtra, India.}

\vspace*{0.225truein}

\publisher{(received date)}{(revised date)}

\setcounter{footnote}{0}
\renewcommand{\thefootnote}{\alph{footnote}}

\vspace*{0.21truein}
\abstracts{The notions of temperature, entropy and `evaporation', usually associated
with spacetimes with horizons, are analyzed using general approach and   the following  results,
 applicable to different 
spacetimes, are obtained  at one go.
 (i) The concept of temperature associated with the horizon is derived in a unified manner and is shown to arise from {\it purely
kinematic considerations}. (ii) QFT near any horizon is mapped to a conformal field theory  without introducing concepts from string theory.  (iii) For  spherically symmetric spacetimes (in $D=1+3$) with a horizon at $r=l$,  the  partition function has the generic form
$Z\propto \exp[S-\beta E]$, where $S= (1/4) 4\pi l^2$  and $|E|=(l/2)$.   This analysis reproduces the conventional result for the blackhole spacetimes and provides a simple and consistent interpretation of entropy and energy for deSitter spacetime. (iv) For the Rindler spacetime the entropy per unit transverse area turns out to be $(1/4)$ while the energy is zero.  (v) In the case
of a Schwarzschild black hole there exist quantum states (like Unruh vacuum)
which are not invariant under time reversal and can describe blackhole evaporation. 
There also exist quantum states (like Hartle-Hawking vacuum) in which  temperature is well-defined  but there is no flow of radiation to infinity. In the case of deSitter universe
or Rindler patch in flat spacetime, one usually uses quantum states analogous to Hartle-Hawking vacuum and obtains a temperature without the corresponding notion
of evaporation. It is, however, possible to construct the analogues of
Unruh vacuum state in the other cases as well. Associating an entropy or a radiating vacuum state with a {\it general} horizon raises conceptual issues
which are briefly discussed.}{}{}

%\vspace*{10pt}
%\keywords{Horizon, Schwarzschild, deSitter, Rindler, Blackhole evaporation, Entropy, Cosmological constant}

\vspace*{1pt}\textlineskip	
\section{Motivation} 	
\vspace*{-0.5pt}
\noindent
One of the remarkable features of classical gravity is that it can wrap up regions of spacetime thereby producing surfaces which act as one way membranes. The classic example is that of a Schwarzschild blackhole which has a compact, observer independent, surface which
acts as an  event horizon.  Another example is the deSitter universe which also has an one way
membrane  though the location of this compact surface depends on the observer.  
 The existence of one-way membranes, however,  is not necessarily a feature of gravity or curved spacetime
and can be induced even in flat Minkowski spacetime, once we accept the notion of  an observer dependent horizon. It is possible to introduce coordinate charts in Minkowski spacetime such that regions are separated by horizons, a familiar example being the Rindler frame which has a non-compact surface acting as a coordinate dependent horizon.

The study of QFT in these  spacetimes suggests a natural way of associating a temperature with the spacetimes which have horizons,  irrespective of whether the horizons are observer independent geometrical structures (as in the case of a blackhole) or observer/coordinate dependent (as in the case of deSitter or Rindler).$^1$  The operator equations for QFT in the background metric are well defined in these spacetimes;  but to make useful predictions we also need to choose a quantum state for the field. The Schwarzschild, deSitter and Rindler metrics are symmetric under time reversal and there exists a `natural' definition of  a time symmetric
vacuum state in all these cases.  Such a vacuum state will appear to be 
described a thermal density matrix in a subregion ${\cal R}$ of spacetime with the horizon as a boundary. The QFT based on such a state will be manifestedly time symmetric and will describe an isolated system in thermal equilibrium in the subregion ${\cal R}$. No time asymmetric phenomena like evaporation, outgoing radiation, irreversible changes etc can take place in this situation. 

One would next ask  whether one can associate an {\it entropy} with such spacetimes in a sensible manner, given that the notion of {\it temperature} arises very naturally. Conventionally there are two  very different ways of defining the entropy, given the notion of temperature: (1)  In statistical mechanics, the   partition function $Z(\beta)$
of the canonical ensemble of systems with constant temperature $\beta^{-1}$  is related to the entropy $S$ and energy $E$ by
$Z(\beta)\propto \exp(S-\beta E)$. (2) In classical thermodynamics, on the other hand, it is the {\it change in} the entropy, which
can be operationally defined via $dS=dE/T(E)$. Integrating this equation will
lead to the function $S(E)$ except for an additive constant which needs to be
determined from additional considerations. Proving the equality of these two concepts was nontrivial and --- historically --- led to the unification of thermodynamics with mechanics. 

In the case of time symmetric vacuum state,  there will be no change of entropy $dS$ and the thermodynamic
route is blocked. I will show, however, that it is possible to construct a canonical
ensemble of a class of  spacetimes with a fixed value for $\beta$ and evaluate the partition function $Z(\beta)$. For  spherically symmetric spacetimes with a horizon at $r=l$,  the  partition function has the generic form
$Z\propto \exp[S-\beta E]$, where $S= (1/4) 4\pi l^2$  and $|E|=(l/2)$. This analysis reproduces the conventional result for the blackhole spacetimes and provides a simple and consistent interpretation of entropy and energy for deSitter spacetime, with the latter being given by $E=-(1/2) H^{-1}$. For the Rindler spacetime the entropy per unit transverse area turns out to be $(1/4)$ while the energy is zero. 

I will also show how to construct radiating states in all these spacetimes such that the thermodynamic approach to entropy can also be realised. It will turn out that 
there is absolutely no mathematical distinction between the 
horizons which arise in the Schwarzschild, deSitter and Rindler spacetimes. There is no simple way one can associate entropy with blackholes {\it without} associating entropy with Rindler or deSitter. It is all or none.

\section{A unified approach to spacetimes with horizons}

\noindent
Consider a (D+1) dimensional flat Lorentzian manifold ${\cal S}$ with the signature (+, $ -,-,  ...)$
and Cartesian coordinates $Z^A$ where $A=(0,1,2,..., D)$. 
A four dimensional sub-manifold ${\cal D}$ in this (D+1) dimensional space can be defined
through a mapping $Z^A=Z^A(x^a)$ where $x^a$ with $a=(0,1,2,3)$ are the four
dimensional coordinates on the surface. The flat Lorentzian metric in the (D+1)
dimensional space induces a metric $g_{ab}(x^a)$ on the four dimensional space
which --- for a wide variety of the mappings $Z^A=Z^A(x^a)$ --- will have the signature
$(+,-,-,-)$ and  will represent, in general, a curved four geometry. The quantum theory of a free scalar field in ${\cal S}$ 
is well defined in terms of the, say, plane wave modes which satisfy the wave equation in ${\cal S}$. A subset of these
modes, which does not depend on the `transverse' directions, will satisfy the corresponding wave equation
in  ${\cal D}$ and will depend only on $x^a$. These modes induce a natural QFT in ${\cal D}$. We are interested
in the mappings $Z^A=Z^A(x^a)$ which leads to a horizon in ${\cal D}$ so that we can investigate the QFT
in spacetimes with horizons using the free, flat spacetime, QFT in ${\cal S}$.  [This approach was used earlier in reference 2 for some special cases.  But
I use this technique with a different motivation, in a more general context, to 
draw some important conclusions.]

 For this purpose, I will restrict attention to a class of surfaces defined by the mappings $Z^A=Z^A(x^a)$ 
which ensures  the following properties for ${\cal D}$: (i) The induced metric $g_{ab}$ has the signature
$(+,-,-,-)$. 
(ii) The induced metric is static in the sense
that $g_{0\alpha}=0$ and all $g_{ab}$s are independent of $x^0$.
[The Greek indices run over 1,2,3.]
  (iii) Under the transformation $x^0 \to x^0+i(\pi/g)$, where
$g$ is a non zero, positive constant, the mapping of the coordinates changes
as $Z^0\to -Z^0, Z^1\to -Z^1$ and $Z^A\to Z^A$ for $A= 2,...,D$. 
It will turn out
that the four dimensional manifolds defined by such mappings possess 
a horizon and most of the interesting features of the thermodynamics related
to the horizon can be obtained from the above characterization. 
Let us first determine the nature of the mapping $Z^A = Z^A(x^a) = Z^A(t,{\bf x})$
such that the above conditions are satisfied. 

The condition (iii) above singles out the spatial coordinate
$Z^1$ from the others. To satisfy this condition we can take the mapping
$Z^A = Z^A(t,r,\theta,\phi)$   to be of the form
$Z^0 =Z^0(t,r), Z^1 =Z^1(t,r), Z^\perp =Z^\perp(r,\theta,\phi)$ where $Z^\perp$
denotes the transverse coordinates $Z^A$ with $A=(2,...,D)$. To impose the condition
(ii) above, one can make use of the fact that ${\cal S}$
 possesses invariance under translations, rotations and Lorentz boosts
which are characterized by the existence of a set of $N = (1/2)(D+1)(D+2)$
Killing vector fields $\xi^A(Z^A)$. Consider any linear combination $V^A$
of these Killing vector fields which is timelike in a region of ${\cal S}$.
 The integral curves to this vector field $V^A$ will
define   time like curves in ${\cal S}$. If one treats these 
curves as the trajectories of a hypothetical observer, then one can
set up the proper Fermi-Walker transported coordinate system
for this observer. Since the four velocity of the observer is along the
Killing vector field, it is obvious that the metric
in this coordinate system will be static.$^3$  In particular,  there
exists a Killing vector which  corresponds to Lorentz boosts along the $Z^1$ direction
that can be interpreted as rotation in imaginary time coordinate allowing a natural realization of (iii) above. 
Using the property of Lorentz boosts, it is easy to see that the transformations
of the form $Z^0 = \pm lf(r)^{1/2} \sinh gt;  Z^1 =  \pm lf(r)^{1/2} \cosh gt$
will satisfy both conditions (ii) and (iii) where $(l,g)$ are constants introduced for dimensional reasons 
and $f(r)$ is a given
 function. This map covers only the two quadrants
with $|Z^1| > |Z^0|$ with positive sign for the right quadrant and negative sign
for the left. To cover the entire $(Z^0,Z^1)$ plane, we will use the full set
\bea
Z^0=\pm lf(r)^{1/2} \sinh gt; \quad Z^1 =  \pm lf(r)^{1/2} \cosh gt \quad &({\rm for}\  |Z^1| >|Z^0|)  \label{eqn:two}
\\
Z^0=\pm l [- f(r)]^{1/2} \cosh gt; \quad Z^1 =   \pm l[- f(r)]^{1/2} \sinh gt \quad &({\rm for}\  |Z^1| <|Z^0|) \label{eqn:twoprime}
\eea
 The inverse
transformations corresponding to (\ref{eqn:two}) are
\bee
l^2f(r)=(Z^1)^2 -(Z^0)^2;\quad gt=\tanh ^{-1}(Z^0/Z^1) \label{eqn:three}
\eeq
Clearly, 
 to cover the entire two dimensional plane of $ - \infty < (Z^0,Z^1) < + \infty $, it is necessary
to have both $f(r)>0$ and $f(r)<0$.  
The pair of points $(Z^0,Z^1)$ and $(-Z^0,-Z^1)$ are mapped to the same
$(t,r)$ making this a 2-to-1 mapping.
The null surface $Z^0=\pm Z^1$ is mapped to the surface $f(r) =0$. 

The transformations given above with any arbitrary mapping for the transverse
coordinate $Z^\perp =Z^\perp(r,\theta,\phi)$ will give rise to an induced 
metric on ${\cal D}$ of the form
\bee
ds^2 = f(r)(lg)^2dt^2 - {l^2 \over 4} \left({ f^{\prime 2} \over f}\right) dr^2 - dL^2_\perp \label{eqn:four}
\eeq
where $dL_\perp^2$ depends on the form of the mapping  $Z^\perp =Z^\perp(r,\theta,\phi)$.
This form of the  metric is valid in all the quadrants even though
we will continue to work in the right quadrant and will comment
on the behaviour in other quadrants only when necessary. It is 
obvious that the ${\cal D}$, in general, is curved and has a horizon
at $f(r)=0$.

As a specific example, let us consider the case of (D+1)=6 with the coordinates
$(Z^0,Z^1,Z^2,Z^3,Z^4,Z^5)=(Z^0,Z^1,Z^2,R,\Theta,\Phi)$ and consider a mapping to
4-dimensional subspace in which: (i) The  $(Z^0,Z^1)$ are mapped to $(t,r)$ as before; (ii) the 
spherical coordinates $(R,\Theta,\Phi)$ in ${\cal S}$ are mapped to standard spherical polar coordinates in
${\cal D}$: $(r, \theta, \varphi)$ {\it and} (iii) we take $Z^2$ to be an arbitrary function of $r$: $Z^2=q(r)$. This leads to the metric
\bee
ds^2 = A(r) dt^2 - B(r) dr^2 - r^2 d \Omega^2_{\rm 2-sphere}; \label{eqn:tena}
\eeq
with
\bee
A(r) = (lg)^2f; \qquad B(r) = 1 + q^{\prime 2} + {l^2 \over 4} {f^{\prime 2 }\over f} \label{eqn:eleven}
\eeq
Equation (\ref{eqn:tena}) is the form of a general, spherically symmetric, static metric in 4-dimension with two arbitrary functions $f(r),q(r)$. Given any specific metric
with $A(r)$ and $B(r)$, equations (\ref{eqn:eleven}) can be solved to determine
$f(r),q(r)$. As an example, let us consider the Schwarzschild solution for which we will take $f=4 \left[ 1 - (l/ r) \right]$; the condition $g_{00}=(1/g_{11})$ now determines $q(r)$ through the equation
\bee
(q^{\prime})^ 2 = \left( 1 + {l^2 \over r^2} \right) \left( 1 + {l \over r} \right) - 1 = \left( {l \over r}\right)^3 + \left({l  \over r}\right)^2 + {l \over r}\label{eqn:twelve}
\eeq
That is
\bee
q(r) = \int^r\left[\left( {l \over r}\right)^3 + \left({l  \over r}\right)^2 + {l \over r}\right]^{1/2} dr \label{eqn:thirteen}
\eeq
Though the integral cannot be expressed in terms of elementary functions, it
is obvious that $q(r)$ is well behaved everywhere including at $r=l$. The transformations 
$(Z^0,Z^1) \to (t,r); Z^2\to q(r); (Z^3,Z^4,Z^5)\to(r, \theta, \varphi)$
thus provide the embedding of Schwarzschild metric in a 6-dimensional space.
[This result was originally obtained by Frondsal;$^4$  but the derivation in that paper is somewhat obscure and does not bring out the generality of the situation]. As a corollary, we may note that this procedure leads to a spherically symmetric Schwarzschild-like metric in arbitrary dimension, with the
2-sphere in (\ref{eqn:tena}) replaced any  $N$-sphere. 

Incidentally, the choice $lg=1,f(r)=[1-(r/l)^2]$ will provide an embedding
of the deSitter spacetime in 6-dimensional space with $Z^2=r, (Z^3,Z^4,Z^5)\to (r,\theta,\phi)$. Of course, in this case, one of the coordinates is actually redundant and we can
achieve the embedding in 5-dimensional space. 
    A still more trivial case is that of Rindler metric  which can be obtained 
    with D=3, $lg=1, f(r)=1+2gr$; in this case, the ``embedding'' is just a reparametrization within four
    dimensional spacetime and --- in this case --- $r$ runs in the range $(-\infty, \infty)$.
    The key point is that the metric in  (\ref{eqn:four}) is fairly generic and can describe
    a host of spacetimes with horizons located at $f=0$.

\section{Dimensional reduction and emergence of CFT near a horizon}
\noindent
To investigate the nature of spacetimes with horizons,
we only need to assume that $f(r)$ vanishes at some $r=l$ 
with $f'(l)$ being finite;
such spacetimes have a horizon at $r=l$. 
Let us consider  the example of a QFT for a self-interacting scalar field with a potential $V(\phi)$, in
a spacetime with the metric of the form 
in equation (\ref{eqn:four}):  
\bee
ds^2 = f(r)(lg)^2dt^2 - {l^2 \over 4} \left({ f^{\prime 2} \over f}\right) dr^2 - g_{\alpha \beta} dx^{\alpha}dx^{\beta}; \quad g_{\alpha \beta} = g_{\alpha \beta} (r, {\bf x}_{\bot})
\eeq
where the line element $g_{\alpha \beta} dx^{\alpha}dx^{\beta}$ denotes the irrelevant transverse part corresponding to the transverse coordinates
${\bf x}_\perp$, {\it as well as} any regular part of the metric
corresponding to $dr^2$. For example, in the case of metric in (\ref{eqn:tena})
it is convenient to combine the $(1+q'^2) dr^2$ term with the transverse part
since it is regular at the horizon.
 The
field equation for a scalar field in this metric $\nabla_a \nabla^a \phi = - (\partial V/\partial \phi)$
can be expanded as
\bee
 \ddot\phi - {4g^2\over Q} {f\over f'} {\partial\over \partial r} \left[ Q {f\over f'} {\partial \phi\over \partial r}\right] - 
  (lg)^2 f \left[ (\nabla^2_{\bot}\phi) + {\partial V \over \partial \phi} \right] = 0\label{eqn:twonine}
\eeq
where we have set $\sqrt{-g} = (l^2g/2) f' (r) Q(r,{\bf x_\perp})$ with $Q(r,{\bf x_\perp})$ giving the square root of the determinant of 
transverse metric. It is assumed that $f$ has a simple zero at some
$r=l$ while $Q$ is regular at this point.
 We see from (\ref{eqn:twonine}) that something drastic will happen if $f$ has a simple zero at some point $r=l$, signalling a horizon in the spacetime. Then the   factor $f$ will kill the dependence of the solution on the transverse coordinates as
well as on the potential and we will  be left with essentially a $2-$ dimensional
scalar field theory, which  -- as is well known -- acquires an extra symmetry of
conformal invariance. 
Introducing the coordinate 
$\xi =  (1 / 2g) \ln f $,
  equation (\ref{eqn:twonine}) can be rewritten as
\bee
{\partial^2 \phi \over \partial t^2} - {\partial^2 \phi \over \partial \xi^2}  = \left( {2g\over f'}\right)^2 {f^2\over Q} \left({\partial \phi\over \partial r}\right) \left({\partial Q\over \partial r}\right) +  (lg)^2 f \left[ (\nabla^2_{\bot}\phi) + {\partial V \over \partial \phi} \right]
\eeq
The right hand side vanishes as $r\to l$ because $f$ vanishes faster than
all other terms. It follows that near the horizon we are dealing with
a (1+1) dimensional field theory governed by
\bee
{\partial^2 \phi \over \partial t^2} - {\partial^2 \phi \over \partial \xi^2}
\approx 0 
\eeq
which has an extra symmetry of 
conformal invariance.

If we take $f(r)\approx 2g (r-l)$ near $r=l$ and separate the time dependence by
$\phi = \phi_\omega e^{-i\omega t}$, it is easy to see that near
$r=l$, the solution  has the universal form:
\bee
\phi_{\omega} \cong \vert r - l \vert^{\pm (i\omega / 2g)} \cong \exp \left[ \pm {i\omega \over 2g} \ln \vert {r \over l} -1 \vert \right] \label{eqn:twentyseven}
\eeq
The fundamental wave modes are now
\bee
\phi = e^{-i\omega t \pm i \omega \xi} = e^{-i \omega (t \pm \xi)}  
 = \left( e^{-i \omega z}, e^{-i\omega \bar z } \right)\label{eqn:thirty}
\eeq
where $\tau = it$ and $z \equiv (\xi + i \tau)$ is the standard complex coordinate of the conformal field theory. The boundary condition on the horizon can be expressed most naturally in terms of $z$ and $\bar z$. For example, purely in-going modes are characterized by $(\partial f/\partial \bar z)=0; (\partial f / \partial z) \not= 0$. Since the system is periodic in $\tau$, the coordinate $z$ is on a cylinder
$(R^1\times S^1$) with $\tau$ being the angular coordinate ($S^1$) and $\xi$ being the $R^1$ coordinate. 
The periodicity in $\tau$ is clearer if we introduce the related complex variable $\rho$ by the definition
$ \rho = \exp g (\xi + i\tau) = \exp(gz) $.
The coordinate $\rho$ respects the periodicity in $\tau$ and is essentially a mapping from a cylinder to a plane.
(For a general discussion of conformal field theory, see reference [5].)
It follows that the modes $(e^{-i \omega z}, e^{-i\omega \bar z})$ become $(\rho^{-i\omega/ g}, \bar\rho^{-i\omega/ g})$ in terms of $\rho$.
 
The situation is simpler in the case of a free field with $V=0$. Then
the general solution to the wave equation can be expanded in terms of the modes
$ \phi (t, r, {\bf x}_{\bot}) = F_{\lambda} ({\bf x}_{\bot} ) \phi_{\lambda \omega}(r) e^{-i \omega t} $
where the function $F$ is the eigenfunction of transverse Laplacian with
(set of) eigenvalue(s) $\lambda$;
that is $ \nabla^2 _{D-1} F=-\lambda^2F$.
In general, the radial part of the solution $\phi_{\lambda\omega}(r)$ will
depend both on $\omega$ and $\lambda$ and will have a complicated $r$ dependence. But if the spacetime has a horizon, then the mode functions have a unique behaviour [given by (\ref{eqn:twentyseven})] near the horizon. 
The corresponding two point function 
$ G\left(t -t^{\prime}; r, r^{\prime} ; {\bf x}_{\bot}, {\bf x}^{\prime}_{\bot} \right) = \langle 0 \vert \phi (t, r, {\bf x}_{\bot} ) \phi (t^{\prime},r^{\prime}, {\bf x}^{\prime}_{\bot})\vert 0 \rangle $
will have the limit 
\bee
G\left(t -t^{\prime}; r, r^{\prime} ; {\bf x}_{\bot} , {\bf x}^{\prime}_{\bot} \right)\cong \left\{ \sum\limits_{\lambda} f_{\lambda}({\bf x}_{\bot}) f_{\lambda}({\bf x}^{\prime}_{\bot} \right\} \left\{ \sum\limits_{\omega} e^{-i \omega[(t-t^{\prime})\pm(\xi - \xi^{\prime})]}\right\} \label{eqn:thirtyseven}
\eeq
near $r \simeq l, r^{\prime} \simeq l$. That is, the two point function factorises into a transverse and radial part
with the radial part being that of a two dimensional massless scalar field. The latter is the same as the Green function of the standard conformal filed theory. Similar results exist for other two-point functions, like Feynman Green function etc.

The fact that the field modes have a unique behaviour near the event horizon
has implications for the QFT in the bulk of the spacetime. Note that a free QFT
is uniquely determined by the two-point function, say, the Feynman Green function which satisfies a local differential equation. Since this equation is hyperbolic in the Minkowski space, the boundary conditions required to specify
the solution are nontrivial in general. Our analysis shows that the existence of a CFT near the horizon allows us to specify the boundary condition on the $r=l+\epsilon$
surface, where $\epsilon$ is an infinitesimal quantity. In other words, the QFT in the bulk of the spacetime is uniquely determined once we specified the boundary conditions on the {\it surface} of the event horizon (plus at infinity or origin depending on the location of bulk region wrt the horizon). This is a version of holographic principle in operation.

    \section{All horizons lead to temperature}
\noindent
There exists a natural definition of QFT in the original (D + 1)-dimensional space; in particular,
we can define a vacuum state for the quantum field on the $Z^0=0$ surface,
which coincides with the $t=0$ surface. By restricting the field modes (or the field configurations
in the Schrodinger picture) to depend only on the coordinates in ${\cal D}$, we will obtain a quantum
field theory in ${\cal D}$ in the sense that these modes will satisfy the relevant field equation
defined in ${\cal D}$. 
In general, this is a complicated problem and it is not easy to have a 
choice of modes in ${\cal S}$ which will lead to a natural set of modes in ${\cal D}$.
We can, however, take advantage of the fact proved in the last section --- that
all the interesting physics arises from the $(Z^0,Z^1)$ plane and the other 
transverse dimensions are irrelevant near the horizon.
In particular, solutions to the wave equation in ${\cal S}$ which depends only on  the 
coordinates $Z^0$ and $Z^1$ will satisfy the wave equation in ${\cal D}$ and will depend
only on $(t,r)$. Such modes will define a natural $s-$wave
QFT in ${\cal D}$.
  I will now show that the positive frequency modes of the above kind will be a specific superposition of
negative and positive frequency modes in ${\cal D}$ leading to
  a  temperature $T=(g/2\pi)$ in the 4-dimensional subspace on
 one side of the horizon. 
There are several ways of proving this result, all of which depend essentially on the property
  that under the transformation $t\to t+(i\pi/g)$ the two coordinates $Z^0$ and $Z^1$ reverses
  sign.

Consider a positive frequency mode of the form
$ F (Z^0,  Z^1) \propto
\exp[-i\Omega Z^0+ iPZ^1]$ with $\Omega > 0$.
This mode
 can be expressed in the 4-dimensional sector in the form $\Phi=F(t,r)= F [Z^0(t,r),  Z^1(t,r)]$.
 For simplicity, we shall measure time in units of $(1/g)$
thereby setting $g=1$ in what follows. The Fourier transform of $F(t,r)$ with respect to $t$ will be:,
\begin{equation}
 K(\omega,r) = \int_{-\infty}^\infty dt\,  e^{-i\omega t} F[Z^0(t,r), Z^1(t,r)]; \qquad (-\infty<\omega<\infty)
 \end{equation}
Thus a positive frequency mode in the higher dimension  can only be
expressed as an integral over $\omega$ with $\omega $ ranging over both positive and
negative values. However, using the fact that $t\to t-i\pi$ leads to $Z^0\to -Z^1$, it is easy to show  that 
$  K(-\omega, r) = e^{-\pi \omega}K^*(\omega,r) $.
This allows us to write
\begin{equation}
F(t,r)=\int_0^\infty {d\omega\over 2\pi}[K(\omega,r)e^{-i\omega t}+
e^{-\pi \omega} K^*(\omega,r)e^{i\omega t}]
\end{equation}
The second term represents the contribution of negative frequency modes in the 
the 4-D spacetime to the pure positive frequency mode in the embedding spacetime.
A field mode of the embedding spacetime containing creation and annihilation operators $(A,A^\dagger)$ can now be
represented in terms of the creation and annihilation operators $(a,a^\dagger)$ appropriate to the
  $(t,r)$ coordinates as 
  \begin{equation}
  AF + A^\dagger F^* = \int_0^\infty {d\omega\over 2\pi} \left[ \left( A + A^\dagger e^{-\pi \omega}\right) Ke^{-i\omega t} + {\rm h.c.}\right] = \int_0^\infty {d\omega\over 2\pi}{1\over N} \left[ a Ke^{-i\omega t} + {\rm h.c.}\right] 
  \end{equation}
  where $N$ is a normalization constant. Identifying $a = N(A+e^{-\pi \omega} A^\dagger)$ and using
  the conditions $[a,a^\dagger] = 1, [A,A^\dagger] = 1$ etc., we get $N=[1-\exp(-2\pi \omega)]^{-1/2}$. It follows that
  the number of $a-$particles in the vacuum defined by $A|{\rm vac}>=0$ is given by 
  \begin{equation}
  <{\rm vac} | a^\dagger a|{\rm vac}> = N^2 e^{-2\pi \omega} = (e^{2\pi \omega}-1)^{-1}
  \end{equation}
  This is a Planckian spectrum with temperature $T=1/2\pi=g/2\pi$ in normal units.

There is a more elegant  way of obtaining this result
which is due to Lee and I give here an adaptation of the same.$^6$ The basic theme is
illustrated in figure 1 in which the origin of $r$ is chosen such that
$f(r=0)=0$ for simplicity.  On the $Z^0=t=0$ hyper-surface
one can define a vacuum state  $|{\rm vac}>$ of the theory by giving the field
configuration for the whole of $-\infty<Z^1<+\infty$. This field configuration, however, separates into two disjoint sectors when one uses
the $(t,r)$ coordinate system.
Concentrating on the $(Z^0,Z^1)$ plane, 
we now need to specify the field configuration
$\phi_R (Z^1)$  for $Z^1>0$ and $\phi_L (Z^1)$ for $Z^1<0$ to match the initial data in
the global coordinates; given this data, the vacuum state is specified by
the  functional $<{\rm vac}|\phi_L,\phi_R>$.

\begin{figure}[htbp] %ORIGINAL SIZE: width=1.4TRUEIN; height=1.5TRUEIN
\vspace*{13pt}
\centerline{\psfig{file=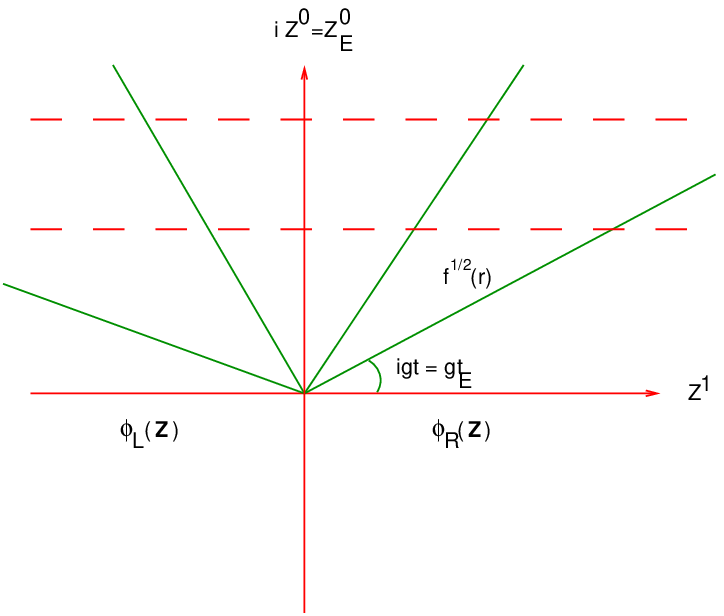}} %100 percent
\vspace*{13pt}
\fcaption{Thermal effects due to a horizon; see text for a discussion.}
\end{figure}

Let us next consider the {\it Euclidean} sector corresponding to the $(Z_E^0,Z^1)$ plane
where $Z_E^0=iZ^0$. The QFT in this plane can be defined along standard lines. The
analytic continuation in $t$, however, is a different matter; the coordinates
$(gt_E=igt,f(r)^{1/2})$ are like polar coordinates in this $(T,Z^1)$ plane with $t_E$ having a periodicity of $(2\pi/g)$. Figure 1 now shows that evolution
in $gt_E$ from $0$ to $\pi$ will take the system configuration from $Z^1>0$ to $Z^1<0$. This allows one to prove that $\langle {\rm vac} |\phi_L , \phi_R \rangle \propto \langle \phi_L \vert e^{-\pi H/g} \vert \phi_R \rangle$; normalization now fixes the proportionality constant, giving 

\bee
\langle {\rm vac} |\phi_L , \phi_R \rangle = {\langle \phi_L \vert e^{-\pi H/g} \vert \phi_R \rangle \over Tr(e^{-2 \pi H/g})^{1/2} } \label{eqn:fifteen}
\eeq
To provide a simple  proof of this relation, let us consider
the ground state wave functional $<{\rm vac} |\phi_L , \phi_R>$ in the embedding spacetime
expressed as a path integral. As is well known, ground state wave functional can be represented
as a Euclidean path integral of the form
\begin{equation}
\langle {\rm vac} |\phi_L , \phi_R \rangle \propto \int_{Z^0_E=0;\phi=(\phi_L,\phi_R)}^{Z^0_E=\infty;\phi=(0,0)}
{\cal D}\phi e^{-A}\label{euclpath}
\end{equation}
  where $Z^0_E=iZ^0 $ is the Euclidean time coordinate. From figure 1 it is obvious that
  this path integral could also be evaluated in the polar coordinates by varying the angle
  $\theta = gt_E$ from 0 to $\pi$. When $\theta =0$ the field configuration corresponds to 
  $\phi = \phi_R$ and when $\theta = \pi$ the field configuration corresponds to $\phi = \phi_L$.
  Therefore
   \begin{equation}
\langle {\rm vac} |\phi_L , \phi_R \rangle \propto \int_{gt_E=0;\phi=\phi_R}^{gt_E=\pi;\phi=\phi_L}
{\cal D}\phi e^{-A}
\end{equation}
   In Heisenberg picture this quantity can be expressed as a matrix element of the Hamiltonian $H_R$
   in the $(t,r)$ coordinates giving us the result: 
   \begin{equation}
\langle {\rm vac} |\phi_L , \phi_R \rangle \propto \int_{gt_E=0;\phi=\phi_R}^{gt_E=\pi;\phi=\phi_L}
{\cal D}\phi e^{-A} = \langle \phi_L|e^{-(\pi/g)H_R}|\phi_R\rangle
\end{equation}
    Normalizing the result properly gives equation (\ref{eqn:fifteen}).

This result, in turn, implies that for operators ${\cal O}$ made out of field variables belonging to the right wedge, the vacuum expectation values become
thermal expectation values. This arises from straightforward algebra of inserting
a complete set of states appropriately:
\bea
&&\langle{\rm vac} \vert \, {\cal O} (\phi_R)\vert {\rm vac}\rangle 
= \sum\limits_{\phi_L} \sum\limits_{\phi^1 _R, \phi^2_R} \langle {\rm vac} \vert \phi_L, \phi^1_R \rangle \langle \phi^1_R \vert {\cal O} (\phi_R)\vert \phi^2_R \rangle \langle \phi^2_R , \phi_L \vert {\rm vac} \rangle \qquad \qquad\quad
 \nonumber \\
&&\quad=\sum\limits_{\phi_L} \sum\limits_{{\phi^1_R}, \phi^2_R} {\langle \phi_L \vert e^{-\pi H/g}\vert \phi^{1}_R \rangle \langle \phi^{1}_R \vert {\cal O} \vert \phi^2_R \rangle \langle \phi^2_R \vert e^{-\pi H/g} \vert \phi_L \rangle \over Tr(e^{-2\pi H/g})}
={Tr (e^{-2\pi H/g} {\cal O}) \over Tr(e^{-2\pi H/g})} \label{eqn:twentytwo}
\eea

The most important conclusion which follows from all these  is that the existence of the temperature is a {\it purely  kinematic effect} arising from the coordinate system we have used.  Dynamical evolution has no role to play.
The main ingredients which have gone into this 
   result are the following. (i) The singular behaviour of the $(t,r)$ coordinate system
   near $r=l$ separates out the $Z^0=0$ hyper-surface into two separate regions.
   (ii) In terms of real $(t,r)$ coordinates, it is not possible to distinguish between the points
   $(Z^0,Z^1)$ and $(-Z^0,-Z^1)$ but the transformation $t\to t+i\pi$ maps the point
   $(Z^0,Z^1)$ to the point $(-Z^0,-Z^1)$. Thus a rotation in the complex time plane
   encodes the information contained in the full $Z^0=0$ plane. 
   
\section{All horizons lead to entropy}
\noindent
The next logical question will be whether one can associate other thermodynamic quantities, especially the entropy, with such spacetimes. Given that the temperature can be introduced very naturally,  using only the behaviour of metric near the horizon, one would look for a similarly elegant and natural derivation of the entropy.
   Such a derivation should not depend on the introduction of external degrees of freedom (like a scalar field) since we want to associate the entropy with the spacetime and not with an external field. Further,  the thermodynamical description should depend only on the behaviour of the metric near the horizon. I will  show that 
  it  is indeed possible to provide such a description  for spacetimes of the form in 
    \begin{equation}
     ds^2=f(r)dt^2-f(r)^{-1}dr^2 -dL_\perp^2
     \label{basemetric}
     \end{equation}
  where $f(r)$ vanishes at some surface $r=l$, say, with $f'(l)\equiv B$ remaining finite. When $dL_\perp^2=r^2dS_2^2$
 with $[0\leq r\leq \infty]$, equation (\ref{basemetric}) covers a variety of spherically symmetric spacetimes  with a compact horizon at $r=l$. If $r$ is interpreted as one of the cartesian coordinates $x$ with $(-\infty\leq x\leq \infty)$ and $dL_\perp^2=dy^2+dz^2, f(x)=1+2gx,$ equation (\ref{basemetric}) can describe the Rindler frame in flat spacetime. We shall first concentrate on compact horizons with $r$ interpreted as radial coordinate,
and comment on the Rindler frame at the end. 

Since the metric is static, Euclidean continuation is trivially effected by $t\to
  \tau=it$ and an examination of the conical singularity near $r=l$ [where $f(r) \approx B(r-l)$] shows that $\tau$ should be interpreted as periodic with period $\beta=4\pi/|B|$ corresponding to the temperature $T=|B|/4\pi$.
  Hence class of metrics in
  (\ref{basemetric}) with the behaviour $[f(l)=0,f'(l)=B]$ constitute a canonical ensemble at constant temperature  since they all have the same temperature $T=|B|/4\pi$ . The partition function for this ensemble is given  by the path integral sum
   \begin{equation}
    Z(\beta)=\sum_{g\epsilon {\cal S}}\exp (-A_E(g))  
 =\sum_{g\epsilon {\cal S}}\exp \left(-{1\over 16\pi}\int_0^\beta  d\tau \int d^3x \sqrt{g_E}R_E[f(r)]\right)
     \label{zdef}
     \end{equation}
  where I have made the Euclidian continuation of the Einstein action and imposed the periodicity in $\tau$
  with period $\beta=4\pi/|B|$.
  The sum is restricted to the set ${\cal S}$ of all metrics of the form in 
  (\ref{basemetric}) with the behaviour $[f(l)=0,f'(l)=B]$ and the Euclidean lagrangian is a functional of $f(r)$.
  No source terms or cosmological constant (which cannot be distinguished from certain form of source) is included since the idea is to obtain a result which depends purely on the geometry.
   The spatial integration will be restricted to a region bounded by the 2-spheres $r=l$ and $r=L$, where
  the choice of $L$ is arbitrary except for the requirement that  within the region of integration the Lorentzian
  metric must have the proper signature with $t$ being a time coordinate.   Using the result
  \begin{equation}
  R=\nabla_r^2 f -{2\over r^2}{d\over dr}\left[r(1-f)\right]\label{reqn}
  \end{equation}
  valid for metrics of the form in (\ref{basemetric}), a
   straight forward calculation shows that
   \begin{equation}
  - A_E={\beta\over 4}\int_l^L dr\left[-[r^2f']'+2[r(1-f)]'\right] 
={\beta\over 4}[l^2B -2l]+Q[f(L),f'(L)]
     \label{zres}
     \end{equation}
where $Q$ depends on the behaviour of the metric near $r=L$ and we have used the 
conditions $[f(l)=0,f'(l)=B]$. The sum in (\ref{zdef}) now reduces to summing over the values of $[f(L),f'(L)]$
with a suitable (but unknown) measure. This sum, however, will only lead to a factor which we can ignore in deciding about the dependence of $Z(\beta)$ on the form of the metric near $r=l$. Using $\beta=4\pi/B$
(and taking $B>0$, for the moment)  the final result can be written in a very suggestive form:
 \begin{equation}
  Z(\beta)=Z_0\exp \left[{1\over 4}(4\pi l^2) -\beta({l\over 2})  \right]\propto 
  \exp \left[S(l) -\beta E(l)  \right]
     \label{zresone}
     \end{equation}
with the identifications for the entropy and energy being given by:
\begin{equation}
S={1\over 4} (4\pi l^2) = {1\over 4} A_{\rm horizon}; \quad E = {1\over 2} l = \left( {A_{\rm horizon}\over 16 \pi}\right)^{1/2}
\end{equation}
  In addition to the simplicity, 
     the following features are noteworthy regarding this result:
   
    (i) The partition function was evaluated with two very natural conditions: $f(l) =0$ making 
    the surface $r=l$ a compact horizon and $f'(l) = $ constant which is the 
    proper characterisation of the canonical ensemble of spacetime metrics. Since temperature
    is well defined for the class of metrics which I have considered, this canonical ensemble is defined without any   ambiguity.  This allows me to  sum over a  class of spherically symmetric spacetimes at one go rather than
    deal with, say, blackhole spacetimes and deSitter spacetime separately. Unlike many of the 
previous approaches (like, for example, ref. [7]),   I do {\it not} evaluate the
path integral in the WKB limit, confining to metrics which are solutions of Einstein's equations. 
 
 (ii) In the case of the Schwarzschild blackhole with $l=2M$, the energy turns out to
    be $E=(l/2) = M$ which is as expected. (More generally,
	$E=(A_{\rm horizon}/16\pi)^{1/2}$ corresponds to the so called `irreducible mass' in 
       BH spacetimes). Of course, the identifications $S=(4\pi M^2)$,
    $E=M$, $T=(1/8\pi M)$ are consistent with the result $dE = TdS$ in this particular case.

   (iii) Most importantly, our analysis provides an interpretation of entropy and energy in the case
    of deSitter universe.  In this case, $f(r) = (1-H^2r^2)$, $a=H^{-1}, B=-2H$.
Since the region where $t$ is timelike is ``inside'' the horizon, the integral for $A_E$ in (\ref{zres}) should be taken from some arbitrary value $r=L$ to $r=l$ with $l>L$. So the horizon contributes in the upper limit of the integral
introducing a change of sign in (\ref{zres}). Further, since $B<0$, there is another negative sign in the area term
from $\beta B\propto B/|B|$. Taking all these into account we get, in this case, 
\begin{equation}
  Z(\beta)=Z_0\exp \left[{1\over 4}(4\pi l^2) +\beta({l\over 2})  \right]\propto 
  \exp \left[S(l) -\beta E(l)  \right]
     \label{zrestwo}
     \end{equation}
giving
$S=(1/ 4) (4\pi l^2) = (1/ 4) A_{\rm horizon}$ and  $E=-(1/2)H^{-1}$. 
These definitions do satisfy the relation $TdS -PdV =dE$ when it is noted that the deSitter universe has 
    a non zero pressure $P=-\rho_\Lambda=-E/V$ associated with the cosmological constant. In fact,
if we use the ``reasonable" assumptions $S=(1/4)(4\pi H^{-2}), V=(4\pi/3)H^{-3}$ and $E=-PV$ in the equation
$TdS -PdV =dE$ and treat $E$ as an unknown function of $H$, we get the equation $H^2(dE/dH)=-(3EH+1)$
which integrates to give precisely $E=-(1/2)H^{-1}. $ This energy is also
is numerically same as the
total energy within the Hubble volume of the classical solution, with a cosmological constant:
    \begin{equation}
    E_{\rm Hub}={4\pi \over 3} H^{-3} \rho_\Lambda = {4\pi \over 3} H^{-3} {3H^2\over 8\pi } ={1\over 2}H^{-1}
    \end{equation}
  (The extra negative sign of $E=-E_{\rm Hub}$ is related to a feature noticed in the literature
in a different context; see for example, the discussion following equation (71) in the review [8]. )
    
Finally, let us consider the spacetimes with planar symmetry for which (\ref{basemetric}) is still applicable with $r=x$ being a Cartesian coordinate and $dL_\perp^2=dy^2+dz^2$. In this case $R=f''(x)$ and the action becomes
\begin{equation}
-A_E={1\over 16\pi}\int_0^\beta d\tau\int dy dz \int_l^L dx f''(x)
={\beta\over 16\pi} A_\perp f'(l)+Q[f'(L)]
\label{rindleraction}
\end{equation}
where we have confined the transverse integrations to a surface of area $A_\perp$. If we now sum over
all the metrics with $f(l)=0,f'(l)=B$ and $f'(L)$ arbitrary, the partition function will become
\begin{equation}
Z(\beta)=Z_0\exp({1\over 4}A_\perp)
\end{equation}
which suggests that planar horizons have an entropy of (1/4) per unit transverse area but zero energy. This includes
Rindler frame as a special case. Note that if we freeze  $f$ to its Rindler form $f=1+2gx$, then $R=f''=0$ as it should.
In the action in (\ref{rindleraction}), $f'(l)-f'(L)$ will give zero. It is only because we are freezing $f'(l)$ but varying $f'(L)$ that we obtain an entropy for these spacetimes.  

\section{QFT in spacetimes with asymptotic horizons}
\noindent
The analysis so far was based on a strictly {\it static} 4-dimensional spacetime obtained as a subspace of the higher dimensional flat manifold. The blackhole metric, for example, corresponds to an eternal blackhole and the vacuum state  which we constructed corresponds to the Hartle-Hawking vacuum of the Schwarzschild spacetime,  describing a blackhole in thermal equilibrium.$^9$ There is no net radiation flowing to infinity and the entropy and temperature obtained in the previous sections were based on equilibrium considerations.

Physically, one would like to have a situation which is asymmetric in time so that one can consider an irreversible flow of energy providing us with a $dE$ that can be used to define a $dS=dE/T$. 
Once  it is realized that only the asymptotic form of the metric matters, we can simplify the above analysis by
just choosing a time {\it asymmetric} vacuum and working with the asymptotic
form of the metric  with the understanding that the asymptotic form arose due to a time asymmetric process (like gravitational collapse).
In the case of blackhole spacetimes this is accomplished --- for example --- by choosing the Unruh vacuum.$^{10}$ The question arises as to how our unified approach fares in handling such a situation which is not time symmetric and the horizon
forms only asymptotically as $t\to\infty$.

There exist analogues of the collapsing blackhole in the case of deSitter and even Rindler. The analogue in the case of deSitter
spacetime will be an FRW universe which behaves like a deSitter universe only at late times.  
Emboldened by the analogy with blackhole spacetime one can also directly construct quantum states (similar to Unruh vacuum of blackhole spacetime) which are time asymmetric, even in the exact deSitter spacetime, with the understanding that the deSitter universe came about at late times through a time asymmetric evolution.

The analogy also works for Rindler spacetime which is also time symmetric.  The coordinate system for an observer with {\it time dependent} acceleration will generalize the  standard Rindler spacetime
in a  time dependent manner. For an observer who was inertial  at early times and is uniformly accelerating at late times, an event horizon forms at late times exactly in analogy with a collapsing blackhole. It is now possible to choose quantum states which are analogous to
Unruh vacuum - that will correspond to an inertial vacuum state at early times and will appear as a thermal state at late times.

It is easy to demonstrate these results in 2D using the
 emergence of CFT near horizon and  calculating expectation values
of the stress tensor especially near the horizon. Concentrating on the $(t,r)$ 
plane, one can introduce a series of natural coordinates of the form 
\begin{equation}
ds^2 = f(r) dt^2 - {dr^2\over f(r)}=f(\xi)(dt^2 - d\xi^2) =f[(v-u)]dudv
\end{equation}
where
\begin{equation}
\xi = \int {dr\over f(r)}; \qquad u= t-\xi; \qquad v=t+\xi
\end{equation}
These coordinate systems are  singular near the horizon $r\approx l$. Assuming
$f(r) \approx B(r - l)$ near $r=l$ [with $B=f'(l)$], it is easy to see that near the horizon
the line element has the form 
$ ds^2\approx Bl e^{(B/2)(v-u)} dv du$.
This form  suggests a natural coordinate transformation from
$(u,v)$ to a non singular coordinate system $(U,V)$ with 
$ V = (2/B) \exp[(B/2)v]; $ $U=-(2/ B) \exp[(-B/2)u]$
in terms of which the metric becomes $ds^2 = -(4f/ B^2 UV)$ $ dUdV$.
By construction, the $(U,V)$ coordinate system is regular on the horizon, since the combination $(f/UV)$ is finite
on the horizon.
(This will lead to the familiar Kruskal coordinate system in the case
of Schwarzschild manifold.) 

Since the mode functions are plane waves in 
conformally flat (1+1) spacetime, we can immediately identify two very 
natural set of modes [and corrsponding vacuum states] in the 
spacetime. The outgoing and ingoing modes of the kind 
$(4\pi \omega)^{-1/2} $ $[\exp(-i\omega u), \exp(-i\omega v)]$
defines a static vacuum state (called Boulware vacuum in the 
case of Schwarzschild blackhole).
The modes of the kind 
$(4\pi \omega)^{-1/2} [\exp(-i\omega U), $ $\exp(-i\omega V)]$
defines another  vacuum state (called Hartle-Hawking vacuum in the 
case of Schwarzschild blackhole). Finally, the modes of the kind
$(4\pi \omega)^{-1/2}  [\exp(-i\omega U), $ $\exp(-i\omega v)]$
 define the analogue of Unruh vacuum. 
 
 In any conformally flat coordinate 
 system of the form $ds^2=C(x^+,x^-)dx^+dx^-$, the expectation values
 of the stress-tensor component are given by$^5$
 \begin{equation}
 <T_{\pm\pm}> = -{1\over 12\pi} C^{1/2} \partial^2_\pm C^{-1/2};\quad
 <T_{+-}> = {C\over 96\pi} R; \quad R=4C^{-1}\partial_+\partial_- \ln C
 \end{equation}
 Using this formula (with $x^\pm$ identified with $(u,v)$ coordinates and the relation
 $\partial_u=(\partial \xi/\partial u) (\partial r/\partial \xi)\partial_r=-(1/2)f(r)\partial_r)$ gives the expectation values 
in the Boulware vacuum: 
 \begin{equation}
 <B|T_{--}|B>=<B|T_{++}|B> = {1\over 96\pi}\left[f f''-{1\over 2}(f')^2\right]
 \end{equation}
wherer $f'=df/dr$ etc.
 Identifying  $x^\pm$  with $(U,V)$ coordinates gives the expectation values 
in the Hartle-Hawking vacuum:
  \begin{equation}
  <HH|T_{--}|HH>=<HH|T_{++}|HH>=<B|T_{--}|B> +  { f'(l)^2\over 192\pi}
  \end{equation}
   In both these cases, there is  no flux since $<T_{rt}>=0$.
  Near the horizon, we have 
  \begin{equation}
  <B|T_{\pm \pm}|B>\approx -{f'(l)^2\over 192\pi};\qquad <HH|T_{\pm \pm}|HH>\approx 0
  \end{equation}
  An inertial observer near the horizon
will use $U$ instead of $u$ and hence the actual values measured by an inertial observer near the horizon will vary as $  
 <B|T_{uu}|B>(du/dU)^2$ and will diverge near the horizon in the Boulware vacuum. 
  
  A more interesting situation arises in the case of Unruh vacuum which differs from the
  Boulware vacuum only in the outgoing modes. If the cordinate $x^-$ is replaced by
  $X^-\equiv F(x^-)$, the conformally flat nature of the line element is maintained and 
  the only stress tensor component which changes is 
  \begin{equation}
  <T_{--}>\;\;\to\;\; <T_{--}>+{1\over 24\pi}
 \left[\left({Q''\over Q}\right)-{1\over 2}\left( {Q'\over Q}\right)^2\right]
 \end{equation}
  where $Q=(1/F')$. Using this we find that
  \begin{equation}
  <U|T_{--}|U>=<HH|T_{--}|HH>; \quad<U|T_{++}|U>=<B|T_{++}|B> 
  \end{equation}
   thereby making
  $<U|T_{--}|U>\ne <U|T_{++}|U> $. This leads to a flux of radiation
  with $<U|T_{r^*t}|U>=-(B^2/192\pi)$. It is also clear that the energy density, as measured by inertial observers, is finite near the horizon.
We thus conclude that we have two acceptable vauum states in {\it all these spacetimes}: (i) the time symmetric Hartle-Hawking state describing thermal equilibrium and zero flux and (ii) the time-asymmetric Unruh vacuum with a flux of radiation.

We cannot, of course, use this trick in 4D away from the horizon and a
 formal analysis of this problem will involve setting up the in and out 
vacua of the theory,  evolving the modes from $t=-\infty$ to $t=+\infty$,
and computing the Bogolibov coefficients. It is, however, not necessary to perform the
details of such an analysis because
all the three spacetimes (Schwarzschild, deSitter and Rindler)
have virtually identical kinematical structure.
In the case of Schwarzschild metric, it is well known that 
the thermal spectrum at late times arises because
the modes
which reach spatial infinity at late times propagate from near the event
horizon at early times and undergo exponential redshift.
The corresponding result occurs in all the 
three spacetimes (and a host of other spacetimes).

Consider the propagation of a wave packet centered around a  radial null ray in a spacetime which - asymptotically - has the form in equation
(\ref{eqn:four}).
The trajectory of the null ray which goes from the initial position $r_{in}$
at $t_{in}$ to a final position $r$ at $t$ is determined by the equation
\bee
t - t_{in} =  \pm\left({1\over 2g}\right)\int_{r_{in}}^r\left({f'\over f}\right)\left (1+\cdots\right)^{1/2}dr\label{eqn:thirtynine}
\eeq
where the $\cdots$ denotes terms arising from the transverse part
containing $dr^2$ (if any),
which are regular at $r=l$. 
Consider now a ray which was close to the horizon initially so that $(r_{in} - l) \ll l$ and propagates to a region far away from the horizon at late times. (In
a black hole metric $r\gg r_{in}$ and the propagation will be outward directed;
in the deSitter metric we will have $r\ll r_{in}$ with rays propagating towards the origin. )
Since we have $f(r) \to 0$ as $r\to l$, the integral will be dominated
by a logarithmic singularity near the horizon and the regular term denoted
by $\cdots$ will not contribute. Then we get
\bee
t - t_{in} =  \pm\left({1\over 2g}\right)\int_{r_{in}}^r\left({f'\over f}\right) \left(1+\cdots\right)^{1/2}dr\approx\pm \left({1\over 2g}\right)\ln |f(r_{in})|+{\rm constant}
\eeq
As the wave propagates away from the horizon its frequency will be redshifted by the factor $\omega \propto (1/ \sqrt{g_{00}})$ so that
\bee
{\omega(t) \over \omega(t_{in})} = \left( {g_{00}(r_{in}) \over g_{00} (r) } \right)^{1/2}
=\left[ {f(r_{\rm in})\over f(r)}\right]^{1/2} \approx K e^{\pm gt}
\label{eqn:fortytwo}
\eeq
where $K$ is an unimportant constant. It is obvious that the dominant behaviour
of $\omega(t)$ will be exponential for any null geodesic starting 
near the horizon and proceeding away since all the transverse
factors will be subdominant to the diverging logarithmic 
singularity arising from the integral of $(1/f(r))$ near the horizon. Thus 
$\omega(t) \propto 
  \exp [\pm gt ]$ 
and the phase $\theta(t)$ of the wave will be vary with time as
$\theta(t) = \int \omega(t)dt  \propto   \exp [\pm gt ]$.
An observer at a fixed  $r$ will see the wave to have the time dependence $\exp  [i \theta (t)]$ which, of course, is not monochromatic. If this wave is decomposed into different Fourier components with respect to $t$, then the amplitude at frequency $\nu$ is given by the Fourier transform
\bee
f(\nu) \propto \int e^{i\theta(t) - i \nu t} dt \propto \int\limits^{\infty}_{-\infty} dt e^{-i(\nu t - Q\exp[\pm gt])}\label{eqn:fortysixa}
\eeq
where $Q$ is some constant.  Changing the variables from $t$ to $\tau$ by  
$ Qe^{\pm gt} = \tau$,
evaluating the integral by analytic continuation to Im $\tau$ and taking the modulus one finds that the result is  a thermal spectrum:
\bee
\vert f (\nu)\vert^2 \propto {1 \over e^{\beta \nu} - 1 } ; \quad \beta = {2\pi \over g}\label{eqn:fortyeight}
\eeq
The standard expressions for the temperature are reproduced for Schwarzschild, deSitter and Rindler spacetimes. 
This analysis  stresses the fact that the origin of thermal spectrum lies in the Fourier transforming of an exponentially redshifted spectrum. 

An interesting question is whether similar results hold for Rindler spacetime  in 4D which, of course, is flat.  
To study this case we need to work with the metric for an observer who is moving with variable acceleration.
The transformation from the flat inertial coordinates $(\bar t,\bar x,\bar y,\bar z)$
to the proper coordinates $(t,x,y,z)$ of an observer with variable acceleration is effected by $\bar y=y, \bar z = z$ and
\bee
\bar x = \int^{\prime} \sinh \Theta (t) dt + x \cosh \Theta (t); \qquad 
\bar t = \int^{\prime} \cosh \Theta (t) dt + x \sinh \Theta (t)\label{eqn:fiftyfive}
\eeq
where the function $\Theta(t)$ is related to the time dependent acceleration
$g(t)$ by
$g(t) = (d \Theta/ dt)$.
The form of the metric in the accelerated frame is remarkably simple:
\bee
ds^2 = (1 +g(t)x)^2 dt^2 - dx^2 - dy^2 - dz^2\label{eqn:fiftyseven}
\eeq
We will treat $g(t)$ to be an arbitrary function except for the limiting behaviour $g(t)\to 0$ for $t\to -\infty$ and $g(t)\to g_0$=constant for $t\to +\infty$. Hence, at early times, the line element in (\ref{eqn:fiftyseven}) represent the standard inertial coordinates and the positive frequency modes define the standard Minkowski vacuum. At late times, the metric goes over to the Rindler coordinates  and we are interested in knowing how the initial vacuum state will be interpreted at late times. The wave equation  $(\Box + m^2)\phi =0$ for a massive scalar field
can be separated in the transverse coordinates as
$ \phi(t,x,y,z)=f(t,x)e^{ik_y y} e^{ik_zz} $
where $f$ satisfies the equation
\bee
- {1 \over (1 +g(t)x)} {\partial \over \partial t} \left( {1 \over (1 + g(t)x)} {\partial f \over \partial x} \right) = \chi^2f \label{eqn:sixty}
\eeq
with $\chi^2\equiv m^2+k_x^2+k_y^2$. It is  possible to solve this partial differential equation 
exactly and obtain the solution
\bee
f(x, t) = f_{k_yk_z \eta}(x, t)
 = \exp -i \chi \left[ \int \cosh (\Theta - \eta) dt + x \sinh (\Theta - \eta) \right] \label{eqn:sixtysix}
\eeq
where $\eta$ is another constant. 
For the limiting behaviour we have assumed for $g(t)$, we see that $\Theta(t)$ vanishes at early times and varies as $\Theta(t)\approx (g_0t+$constant) at late times. Correspondingly, the mode $f$ will behave as
\bee
f(x, t)
\rightarrow \exp -i \chi \left[ t \cosh (\eta)  - x \sinh (\eta) \right]  \label{eqn:sixtyseven}
\eeq
at early times $(t\to-\infty)$ which is just the standard Minkowski positive frequency mode with
$\omega=\chi \cosh\eta, k_x=\chi\sinh\eta$.  At late times the mode evolves to
\bee
f(x, t)\rightarrow \exp -i\left[ (\chi/2g_0)  (1+g_0x)e^{g_0t}\right] \label{eqn:sixtyeight}
\eeq
We are once again led to a wave mode with exponential redshift at any given $x$. The metric
is static in $t$ at late times and the out-vacuum will be defined in terms of
modes which are positive frequency with respect to $t$. The bogolibov transformations between the mode in (\ref{eqn:sixtyeight}) and modes which vary as $\exp(-i\nu t)$
will involve exactly the same mathematics as in  equation (\ref{eqn:fortysixa}). We will get a thermal spectrum at late times.

How do we interpret these results? In the case of Rindler spacetimes, the detection of any particle
has always been attributed to the energy supplied by the source which is providing the 
acceleration --- either constant or variable. The other two examples --- the Schwarzschild and 
the deSitter --- are more non trivial to interpret.
In the case of blackholes, one considers the collapse scenario as ``physical" and the natural quantum state is the Unruh vacuum. The notions of evaporation, entropy etc. then follow in a concrete manner. The eternal blackhole (and the Hartle-Hawking vacuum state) is taken to be just mathematical constructs not realized in nature. 
In deSitter  spacetime there is {\it no}
natural notion of energy (unlike in blackhole spacetimes which are asymptotically flat) or an ``energy source" analogous to the mass of the blackhole. 
If the spacetime is asymptotically deSitter, should one interpret it as ``evaporating" at late times  with the cosmological constant changing with time ? 

 It may seem correct to conclude that the horizons
{\it always} have temperature but it may not be conceptually straight forward to associate an entropy with the horizon in all cases.
  Unfortunately, there is no clear mathematical reason for such a dichotomous approach since:
  (i) The temperature and entropy 
  for  these spacetimes arise in identical manner due to identical mathematical
  formalism.  It will be surprising if one has
  entropy while the other does not. 
  (ii) Just as collapsing blackhole leads to an asymptotic event horizon, a universe
  which is dominated by cosmological constant at late times will also lead to a horizon.
  Just as we can mimic the time dependent effects in a collapsing blackhole by
  a time asymmetric quantum state (say, Unruh vacuum), we can mimic the late time behaviour of 
  an asymptotically deSitter universe by a corresponding time asymmetric quantum state.
  Both these states will lead to stress tensor expectation values in which there will be a flux
  of radiation. 
  One is almost forced to the conclusion that in such a scenario the cosmological constant is evaporating.

\nonumsection{Acknowledgments}
\noindent
I thank Apoorva Patel for useful discussions.

\nonumsection{References}

\end{document}